# Quantum Query Algorithm Constructions for Computing *AND*, *OR* and *MAJORITY* Boolean Functions

Alina Vasiljeva[*]

Institute of Mathematics and Computer Science University of Latvia
Raiņa bulvāris 29, Riga, LV-1459, Latvia
Alina.Vasiljeva@gmail.com

**Abstract.** Quantum algorithms can be analyzed in a query model to compute Boolean functions where input is given in a black box and the aim is to compute function value for arbitrary input using as few queries as possible. We concentrate on quantum query algorithm designing tasks in this paper. The main aim of the research was to find new efficient algorithms and develop general algorithm designing techniques. First, we present several exact quantum query algorithms for certain problems that are better than classical counterparts. Next, we introduce algorithm transformation methods that allow significant enlarging of exactly computable functions sets. Finally, we propose quantum algorithm designing methods. Given algorithms for the set of sub-functions, our methods use them to design a more complex one, based on algorithms described before. Methods are applicable for input algorithms with specific properties and preserve acceptable error probability and number of queries. Methods offer constructions for computing *AND*, *OR* and *MAJORITY* kinds of Boolean functions.

## 1 Introduction

Let $f(x_1, x_2, ..., x_n) : \{0,1\}^n \to \{0,1\}$ be a Boolean function. We have studied the query model, where a black box contains the input $(x_1, x_2, ..., x_n)$ and can be accessed by questioning $x_i$ values. The goal here is to compute the value of the function. The complexity of a query algorithm is measured by number of questions it asks. The classical version of this model is known as *decision trees* [1]. Quantum query algorithms can solve certain problems faster than classical algorithms. The best-known exact quantum algorithm was designed for *PARITY* function with $n/2$ questions vs. $n$ questions required by classical algorithm [2,3].

The problem of quantum algorithm construction is not that easy. Although there is a large amount of lower and upper bound estimations of quantum algorithm complexity [2, 6, 7], examples of non-trivial and original quantum query algorithms are very few. Moreover, there is no special technique described to build a quantum algorithm for a certain function with complexity defined in advance.

[*] Research supported by the European Social Fund

Most probably it would take a lot of time even for experienced quantum computation specialist to construct an efficient query algorithm, for example, for such functions:

$$F_4(x_1, x_2, x_3, x_4) = \neg(x_1 \oplus x_2) \wedge \neg(x_3 \oplus x_4)$$

$$F_6(X) = (\neg(x_1 \oplus x_2) \wedge \neg(x_2 \oplus x_3)) \wedge (\neg(x_4 \oplus x_5) \wedge \neg(x_5 \oplus x_6))$$

or

$$F_{10}(X) = (f_1 \wedge f_2 \wedge f_3) \vee (f_1 \wedge f_2 \wedge f_4) \vee (f_1 \wedge f_3 \wedge f_4) \vee (f_2 \wedge f_3 \wedge f_4), \text{where}$$
$$f_1 = (x_1 \oplus x_2) \vee (x_3 \oplus x_4);\ f_2 = x_5 \oplus x_6;\ f_3 = \neg(x_7 \oplus x_8) \wedge \neg(x_8 \oplus x_9);\ f_4 = \neg x_{10}$$

In our work we have tried to develop general constructions and approaches for computing Boolean functions in quantum query settings.

Boolean functions are widely adopted in real life processes, that is the reason why our capacity to build a quantum algorithm for an arbitrary function appears to be extremely important. While working on common techniques, we are trying to collect examples of efficient quantum algorithms to build up a base for powerful computation using the advantages of the quantum computer.

Paper is organized as follows. Section 2 consists of theoretical background and definitions. In section 3 two exact quantum query algorithm are presented, which will be used as a base in further sections. In section 4 we present three algorithm transformation methods. Section 5 contains the major part of results - algorithm constructions for computing *AND*, *OR* and *MAJORITY* kinds of Boolean functions. Finally, the summary of results is given in section 6.

## 2   Notation and Definitions

Let $f(x_1, x_2, ..., x_n):\{0,1\}^n \rightarrow \{0,1\}$ be a Boolean function. We use $\oplus$ to denote XOR operation (exclusive OR). We use $\bar{f}$ for the function 1 - *f*. We also use abbreviation QQA for "quantum query algorithm".

### 2.1   Quantum computing

We apply the basic model of quantum computing. For more details see textbooks by Gruska [4] and Nielsen and Chuang [5].

An *n*-dimensional quantum pure state is a vector $|\psi\rangle \in C^n$ of norm 1. Let $|0\rangle, |1\rangle, ...,$ $|n\text{-}1\rangle$ be an orthonormal basis for $C^n$. Then, any state can be expressed as $|\psi\rangle = \sum_{i=0}^{n-1} a_i |i\rangle$ for some $a_i \in C$. Since the norm of $|\psi\rangle$ is 1, we have $\sum_{i=0}^{n-1} |a_i|^2 = 1$. States $|0\rangle, |1\rangle, ..., |n\text{-}1\rangle$ are called *basic states*. Any state of the form $\sum_{i=0}^{n-1} a_i |i\rangle$ is called a *superposition* of $|0\rangle, ..., |n\text{-}1\rangle$. The coefficient $a_i$ is called an *amplitude* of $|i\rangle$.

The state of a system can be changed using *unitary transformations.* Unitary transformation $U$ is a linear transformation on $C^n$ that maps vector of unit norm to vectors of unit norm.

The simplest case of quantum measurement is used in our model. It is the full measurement in the computation basis. Performing this measurement on a state $|\psi\rangle = a_0|0\rangle + \ldots a_k|k\rangle$ gives the outcome $i$ with probability $|a_i|^2$. The measurement changes the state of the system to $|i\rangle$ and destroys the original state $|\psi\rangle$.

### 2.2 Query model

Query algorithm is a model for computing Boolean functions. In this model, a black box contains the input $(x_1, x_2, \ldots, x_n)$ and can be accessed by questioning $x_i$ values. Query algorithm must be able to determine the value of a function correctly for arbitrary input contained in a black box. The complexity of the algorithm is measured by the number of queries to the black box which it uses. The classical version of this model is known as *decision trees*. For details, see the survey by Buhrman and de Wolf [1].

We consider computing Boolean functions in the quantum query model. For more details, see the survey by Ambainis [6] and textbooks by Gruska [4] and de Wolf [2]. A quantum computation with $T$ queries is a sequence of unitary transformations:

$$U_0 \to Q_0 \to U_1 \to Q_1 \to \ldots \to U_T \to Q_{T-1} \to U_T$$

$U_i$'s can be arbitrary unitary transformations that do not depend on the input bits $x_1, x_2, \ldots, x_n$. $Q_i$'s are query transformations. Computation starts in the state $|\vec{0}\rangle$. Then we apply $U_0, Q_0, \ldots, Q_{T-1}, U_T$ and measure the final state.

There are several different, but equally acceptable ways to define quantum query algorithms [2]. The most important consideration is to choose an appropriate definition for the query black box, defining a way of asking questions and receiving answers from the oracle.

Next we will precisely describe the full process of quantum query algorithm definition and notation used in this paper.

Each quantum query algorithm is characterized by the following parameters:

1) *Unitary transformations*
All unitary transformations and the sequence of their application (including the query transformation parts) should be specified. Each unitary transformation is a unitary matrix.

Here is an example of an algorithm sequence specification with $T$ queries:

$$|\vec{0}\rangle \to U_0 \to Q_1 \to \ldots \to Q_T \to U_N \to [QM],$$

where is initial state, [QM] – quantum measurement.

For convenience we will use *bra* notation for describing state vectors and algorithm flows. Quantum mechanics employs the following notation for state vectors [5]:

$$\text{Ket notation: } |\psi\rangle = \begin{pmatrix} \alpha_1 \\ ... \\ \alpha_n \end{pmatrix} \quad \text{Bra notation: } \langle\psi| = |\psi\rangle^+ = (\alpha_1^*, \ ..., \ \alpha_n^*)$$

Algorithm designed in *bra* notation can be converted to *ket* notation by replacing each unitary transformation matrix with its adjoint matrix (conjugate transpose):

Quantum query algorithm flow in *bra* notation: $\langle\psi| = \langle\vec{0}|U_0 Q_0 ... Q_{N-1} U_N$

Quantum query algorithm flow in *ket* notation: $|\psi\rangle = U_N^+ Q_{N-1}^+ ... Q_0^+ U_0^+ |\vec{0}\rangle$

2) *Queries*
We use the following definition of query transformation: if input is a state $|\psi\rangle = \sum_i a_i |i\rangle$, then the output is $|\phi\rangle = \sum_i (-1)^{x_k} a_i |i\rangle$, where we can arbitrary choose variable assignment $x_k$ for each amplitude $\alpha_i$. Assume we have a quantum state with *m* amplitudes $\langle\psi| = (\alpha_1, \alpha_2, ..., \alpha_m)$. For the *n* argument function, we define a query as $QQ_i = (\alpha_1 \equiv k_1, ..., \alpha_m \equiv k_m)$, where *i* is the number of question and $k_j \in \{1..n\}$ is the number of queried variable for *j*-th amplitude (*QQ* abbreviates "quantum query"). If $x_{k_j} = 1$, a query will change the sign of the *j*-th amplitude to the opposite sign; in other case, the sign will remain as-is. Unitary matrix that corresponds to query transformation $QQ_i = (\alpha_1 \equiv k_1, ..., \alpha_m \equiv k_m)$ is:

$$QQ_i = \begin{pmatrix} (-1)^{X_{k1}} & 0 & ... & 0 \\ 0 & (-1)^{X_{k2}} & ... & 0 \\ ... & ... & ... & ... \\ 0 & 0 & ... & (-1)^{X_{km}} \end{pmatrix}$$

3) *Measurement*
Each basic state of a quantum system corresponds to the algorithm output. We assign a value of a function to each output. We denote it as $QM = (\alpha_1 \equiv k_1, ..., \alpha_m \equiv k_m)$, where $k_i \in \{0,1\}$ (*QM* abbreviates "quantum measurement"). The result of running algorithm on input *X* is *j* with a probability that equals the sum of squares of all amplitudes, which corresponds to outputs with value *j*.

Very convenient way of quantum query algorithm representation is a graphical picture and we will use this style when describing designed quantum query algorithms.

### 2.3 Query Algorithm Complexity

The complexity of a query algorithm is based on the number of questions it uses to determine the value of a function on worst-case input.

The *deterministic complexity* of a function $f$, denoted by $D(f)$, is the maximum number of questions that must be asked on any input by a deterministic algorithm for $f$ [1].

The sensitivity of $f$ on input $(x_1, x_2, \ldots, x_n)$ is the number of variables $x_i$ with the following property: $f(x_1, \ldots, x_i, \ldots, x_n) \neq f(x_1, \ldots, 1-x_i, \ldots, x_n)$. The sensitivity of $f$ is the maximum sensitivity of all possible inputs. It has been proved that $D(f) \geq s(f)$ [1].

A quantum query algorithm *computes f exactly* if the output equals $f(x)$ with a probability 1, for all $x \in \{0,1\}^n$. Complexity is denoted by $Q_E(f)$ [1].

A quantum query algorithm *computes f with bounded-error* if the output equals $f(x)$ with probability $p > 1/2$, for all $x \in \{0,1\}^n$. Complexity is denoted by $Q_P(f)$ [1].

## 3 Basic Exact Quantum Query Algorithms

In this section we present two basic exact quantum query algorithms, which will be used as a base for construction methods in further sections.
First algorithm computes 3-argument Boolean function, but second one computes 4-argument Boolean function. Both algorithms are interesting first of all because they are better than best possible classical algorithms. Secondly, algorithms satisfy specific properties, which make them useful for computing more complex Boolean functions.

### 3.1 3-variable function with 2 queries

In this section we present quantum query algorithm for 3-variable Boolean function that saves one query comparing to the best possible classical deterministic algorithm.

**Problem:** *Check if all input variable values are equal.*

Possible real life application is, for example, automated voting system, where statement is automatically approved only if all participants voted for acceptance/rejection equally. We provide solution for 3-party voting routine. We reduce a problem to computing the following Boolean function defined by the logical formula: $EQUALITY_3(X) = \neg(x_1 \oplus x_2) \wedge \neg(x_2 \oplus x_3)$.

**Deterministic complexity:** $D(EQUALITY_3)=3$, by sensitivity on any accepting input.

**Algorithm 1.** Exact quantum query algorithm for $EQUALITY_3$ is presented in figure 1. Each horizontal line corresponds to the amplitude of the basic state. Computation starts with amplitude distribution $\langle \vec{0}| = (1,0,0,0)$. Three large rectangles correspond to the 4x4 unitary matrices ($U_0$, $U_1$, $U_2$). Two vertical layers of circles specify the queried variable order for each query ($Q_0$, $Q_1$). Finally, four small squares at the end of each horizontal line define the assigned function value for each output.

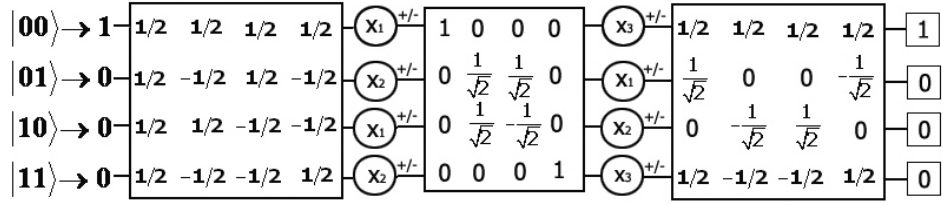

Fig. 1. Exact quantum query algorithm for *EQUALITY*$_3$

We show the computation process for accepting input X=111:

$$\langle\psi| = (1/2,\ 1/2,\ 1/2,\ 1/2)Q_0 U_1 Q_1 U_2 = (-1/2, -1/2, -1/2, -1/2) U_1 Q_1 U_2 =$$

$$= (-1/2,\ -1/\sqrt{2},\ 0,\ -1/2) Q_1 U_2 = (1/2,\ 1/\sqrt{2},\ 0,\ 1/2) U_2 = \mathbf{(1,0,0,0)} \Rightarrow [\text{ACCEPT}]$$

Table 1 shows computation process for each possible input. Processing result always equals *EQUALITY*$_3$ value with probability *p*=1.

Table 1. Quantum query algorithm computation process for *EQUALITY*$_3$

| X | after $\langle\vec{0}|U_0 Q_0$ | after $\langle\vec{0}|U_0 Q_0 U_1 Q_1$ | final state | result |
|---|---|---|---|---|
| 000 | $\left(\frac{1}{2},\frac{1}{2},\frac{1}{2},\frac{1}{2}\right)$ | $\left(\frac{1}{2},\frac{1}{\sqrt{2}},0,\frac{1}{2}\right)$ | (1,0,0,0) | **1** |
| 001 | $\left(\frac{1}{2},\frac{1}{2},\frac{1}{2},\frac{1}{2}\right)$ | $\left(-\frac{1}{2},\frac{1}{\sqrt{2}},0,-\frac{1}{2}\right)$ | (0,0,0,-1) | **0** |
| 010 | $\left(\frac{1}{2},-\frac{1}{2},\frac{1}{2},-\frac{1}{2}\right)$ | $\left(\frac{1}{2},0,\frac{1}{\sqrt{2}},-\frac{1}{2}\right)$ | (0,0,1,0) | **0** |
| 011 | $\left(\frac{1}{2},-\frac{1}{2},\frac{1}{2},-\frac{1}{2}\right)$ | $\left(-\frac{1}{2},0,\frac{1}{\sqrt{2}},\frac{1}{2}\right)$ | (0,-1,0,0) | **0** |
| 100 | $\left(-\frac{1}{2},\frac{1}{2},-\frac{1}{2},\frac{1}{2}\right)$ | $\left(-\frac{1}{2},0,\frac{1}{\sqrt{2}},\frac{1}{2}\right)$ | (0,-1,0,0) | **0** |
| 101 | $\left(-\frac{1}{2},\frac{1}{2},-\frac{1}{2},\frac{1}{2}\right)$ | $\left(\frac{1}{2},0,\frac{1}{\sqrt{2}},-\frac{1}{2}\right)$ | (0,0,1,0) | **0** |
| 110 | $\left(-\frac{1}{2},-\frac{1}{2},-\frac{1}{2},-\frac{1}{2}\right)$ | $\left(-\frac{1}{2},\frac{1}{\sqrt{2}},0,-\frac{1}{2}\right)$ | (0,0,0,-1) | **0** |
| 111 | $\left(-\frac{1}{2},-\frac{1}{2},-\frac{1}{2},-\frac{1}{2}\right)$ | $\left(\frac{1}{2},\frac{1}{\sqrt{2}},0,\frac{1}{2}\right)$ | (1,0,0,0) | **1** |

## 3.2 4-variable function with 2 queries

In this section we present our solution for the computational problem of comparing elements of a binary string.

**Problem:** *For a binary string of length 2k check if elements are equal by pairs:*

$$x_1=x_2,\ x_3=x_4,\ x_5=x_6,...,\ x_{2k-1}=x_{2k}$$

We present an algorithm for string of length 4. We reduce the problem to computing the Boolean function of 4 variables. Boolean function can be represented by formula:

$$PAIR\_EQUALITY_4(x_1, x_2, x_3, x_4) = \neg(x_1 \oplus x_2) \wedge \neg(x_3 \oplus x_4).$$

**Deterministic complexity**: $D(PAIR\_EQUALITY_4)=4$, by sensitivity on accepting input.

**Algorithm 2**. Exact quantum query algorithm for $PAIR\_EQUALITY_4$ is presented in figure 2.

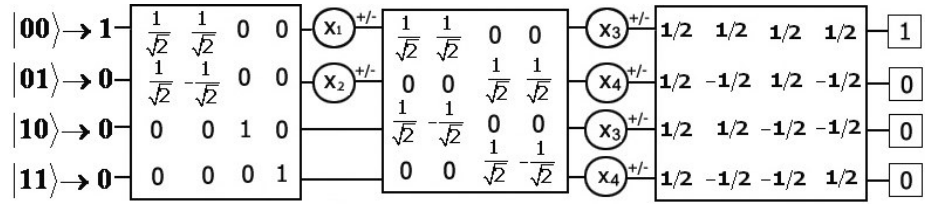

Fig. 2 Exact quantum query algorithm for $PAIR\_EQUALITY_4$

Computational flow for each function input is presented in table 2.

Table 2. Quantum query algorithm computation process for $PAIR\_EQUALITY_4$

| X | after $\langle \vec{0} | U_0 Q_0$ | after $\langle \vec{0} | U_0 Q_0 U_1 Q_1$ | final state | result |
|---|---|---|---|---|
| 0000 | $\left(\frac{1}{\sqrt{2}}, \frac{1}{\sqrt{2}}, 0, 0\right)$ | $\left(\frac{1}{2}, \frac{1}{2}, \frac{1}{2}, \frac{1}{2}\right)$ | (1,0,0,0) | **1** |
| 0001 | $\left(\frac{1}{\sqrt{2}}, \frac{1}{\sqrt{2}}, 0, 0\right)$ | $\left(\frac{1}{2}, -\frac{1}{2}, \frac{1}{2}, -\frac{1}{2}\right)$ | (0,1,0,0) | **0** |
| 0010 | $\left(\frac{1}{\sqrt{2}}, \frac{1}{\sqrt{2}}, 0, 0\right)$ | $\left(-\frac{1}{2}, \frac{1}{2}, -\frac{1}{2}, \frac{1}{2}\right)$ | (0,-1,0,0) | **0** |
| 0011 | $\left(\frac{1}{\sqrt{2}}, \frac{1}{\sqrt{2}}, 0, 0\right)$ | $\left(-\frac{1}{2}, -\frac{1}{2}, -\frac{1}{2}, -\frac{1}{2}\right)$ | (-1,0,0,0) | **1** |
| 0100 | $\left(\frac{1}{\sqrt{2}},-\frac{1}{\sqrt{2}},0,0\right)$ | $\left(\frac{1}{2},\frac{1}{2},-\frac{1}{2},-\frac{1}{2}\right)$ | (0,0,1,0) | **0** |
| 0101 | $\left(\frac{1}{\sqrt{2}},-\frac{1}{\sqrt{2}},0,0\right)$ | $\left(\frac{1}{2},-\frac{1}{2},-\frac{1}{2},\frac{1}{2}\right)$ | (0,0,0,1) | **0** |
| 0110 | $\left(\frac{1}{\sqrt{2}},-\frac{1}{\sqrt{2}},0,0\right)$ | $\left(-\frac{1}{2},\frac{1}{2},\frac{1}{2},-\frac{1}{2}\right)$ | (0,0,0,-1) | **0** |
| 0111 | $\left(\frac{1}{\sqrt{2}},-\frac{1}{\sqrt{2}},0,0\right)$ | $\left(-\frac{1}{2},-\frac{1}{2},\frac{1}{2},\frac{1}{2}\right)$ | (0,0,-1,0) | **0** |
| 1000 | $\left(-\frac{1}{\sqrt{2}},\frac{1}{\sqrt{2}},0,0\right)$ | $\left(-\frac{1}{2},-\frac{1}{2},\frac{1}{2},\frac{1}{2}\right)$ | (0,0,-1,0) | **0** |
| 1001 | $\left(-\frac{1}{\sqrt{2}},\frac{1}{\sqrt{2}},0,0\right)$ | $\left(-\frac{1}{2},\frac{1}{2},\frac{1}{2},-\frac{1}{2}\right)$ | (0,0,0,-1) | **0** |
| 1010 | $\left(-\frac{1}{\sqrt{2}},\frac{1}{\sqrt{2}},0,0\right)$ | $\left(\frac{1}{2},-\frac{1}{2},-\frac{1}{2},\frac{1}{2}\right)$ | (0,0,0,1) | **0** |
| 1011 | $\left(-\frac{1}{\sqrt{2}},\frac{1}{\sqrt{2}},0,0\right)$ | $\left(\frac{1}{2},\frac{1}{2},-\frac{1}{2},-\frac{1}{2}\right)$ | (0,0,1,0) | **0** |
| 1100 | $\left(-\frac{1}{\sqrt{2}},-\frac{1}{\sqrt{2}},0,0\right)$ | $\left(-\frac{1}{2},-\frac{1}{2},-\frac{1}{2},-\frac{1}{2}\right)$ | (-1,0,0,0) | **1** |
| 1101 | $\left(-\frac{1}{\sqrt{2}},-\frac{1}{\sqrt{2}},0,0\right)$ | $\left(-\frac{1}{2},\frac{1}{2},-\frac{1}{2},\frac{1}{2}\right)$ | (0,-1,0,0) | **0** |
| 1110 | $\left(-\frac{1}{\sqrt{2}},-\frac{1}{\sqrt{2}},0,0\right)$ | $\left(\frac{1}{2},-\frac{1}{2},\frac{1}{2},-\frac{1}{2}\right)$ | (0,1,0,0) | **0** |
| 1111 | $\left(-\frac{1}{\sqrt{2}},-\frac{1}{\sqrt{2}},0,0\right)$ | $\left(\frac{1}{2},\frac{1}{2},\frac{1}{2},\frac{1}{2}\right)$ | (1,0,0,0) | **1** |

## 4  Algorithm Transformation Methods

In this section we introduce quantum query algorithm transformation methods that can be useful for enlarging a set of exactly computable Boolean functions. Each method receives exact QQA on input, processes it as defined, and as a result slightly different exact algorithm is obtained that computes another function.

### 4.1  Output value assignment inversion

The first method is the simplest one. All we need to do with original algorithm is to change assigned function value for each output to the opposite.

| *First transformation method - Output value assignment inversion* |
|---|

**Input.** An arbitrary exact QQA that computes *f(X)*.

**Transformation actions.**
- For each algorithm output change assigned value of function to opposite.
  If original assignment was $QM = (\alpha_1 \equiv k_1,...,\alpha_m \equiv k_m)$, where $k_i \in \{0,1\}$,
  Then it is transformed to $QM' = (\alpha_1 \equiv \bar{k}_1,...,\alpha_m \equiv \bar{k}_m)$, where $\bar{k}_i = 1 - k_i$.

**Output.** An exact QQA that computes $\bar{f}(X)$.

Box 1. Description of the first transformation method

### 4.2  Output value assignment permutation

Describing next method we will limit ourselves to using only exact QQA with specific properties as an input for transformation method.

**Property 1.** *We say that exact QQA satisfies Property 1 IFF on any input system state before a measurement is such that for exactly one amplitude $\alpha_i$ holds true that* $|\alpha_i|^2 = 1$. *For other amplitudes holds true that* $|\alpha_j|^2 = 0$, *for* $\forall j \neq i$.

Algorithm 1 and Algorithm 2 from section 3 satisfy Property 1.

| *Second transformation method - Output value assignment permutation* |
|---|

**Input.**
- An exact QQA satisfying *Property 1* that computes *f(X)*.
- Permutation $\sigma$ of the set $OutputValues = \{k_1, k_2, ..., k_m\}$.

**Transformation actions.**
- Permute function values assigned to outputs in order specified by $\sigma$.
  If original assignment was $QM = (\alpha_1 \equiv k_1,...,\alpha_m \equiv k_m)$, where $k_i \in \{0,1\}$,
  Then it is transformed to $QM' = (\alpha_1 \equiv \sigma(k_1),...,\alpha_m \equiv \sigma(k_m))$.

**Output.** An exact QQA for some function *g(X)*.

Box 2. Description of the second transformation method

**Proof of correctness.** Application of the method doesn't break the exactness of QQA, because the essence of *Property 1* is that before the measurement we always obtain non-zero amplitude in exactly one output. Since function value is clearly specified for each output we would always observe specific value with probability 1 for any input. □

The structure of new function *g(X)* strictly depends on internal properties of original algorithm. To explicitly define new function one needs to inspect original algorithm behavior on each input and construct a truth table for new output value assignment.

### 4.3 Query variable permutation

Let $\sigma$ be a permutation of the set $\{1, 2, ..., n\}$, where elements correspond to variable numbers. By saying that function *g(X)* is obtained by permutation of *f(X)* variables we mean the following: $g(X) = f\left(x_{\sigma(1)}, x_{\sigma(2)}, ..., x_{\sigma(n)}\right)$. In our third transformation method we expand the idea of variable permutation to QQA algorithm definition.

---

*Third transformation method – Query variable permutation*

**Input.**
- An arbitrary exact QQA that computes $f_n(X)$.
- Permutation $\sigma$ of variable numbers $VarNum = \{0, 1, ..., n\}$.

**Transformation actions.**
- Apply permutation of variable numbers $\sigma$ to all query transformations.
  If original *i*-th query was defined as $QQ_i = (\alpha_1 \equiv k_1, ..., \alpha_m \equiv k_m)$,
  Then it is transformed to $QQ_i' = (\alpha_1 \equiv \sigma(k_1), ..., \alpha_m \equiv \sigma(k_m))$, $k_i \in \{1, ..., n\}$.

**Output.** An exact QQA computing a function $g(X) = f\left(x_{\sigma(1)}, x_{\sigma(2)}, ..., x_{\sigma(n)}\right)$.

---

Box 3. Description of the third transformation method

**Proof of correctness.** If we apply transformation method described in Box 3, variable values will influence new algorithm flow according to the order specified by permutation $\sigma$, thus an algorithm computes *g(X)* instead of *f(X)*. □

### 4.4 Results of Applying Transformation Methods

Now we will demonstrate transformation methods application results for basic exact algorithms from section 3.

By using $EQUALITY_3$ function we obtained a set of 3-argument Boolean functions, we denote it with *QFunc3*, where for each function there is an exact QQA which computes it with 2 queries. Totally 8 different functions were obtained $|QFunc3| = 8$. Functions are presented in table 3.

Table 3. Results of applying transformation methods for $EQUALITY_3$ algorithm (set $QFunc3$)

| X | EQUALITY | Output value assignment pernutation | | | Output value assignment inversion | | | |
|---|---|---|---|---|---|---|---|---|
| | (1,0,0,0) | (0,1,0,0) | (0,0,1,0) | (0,0,0,1) | (0,1,1,1) | (1,0,1,1) | (1,1,0,1) | (1,1,1,0) |
| 000 | 1 | 0 | 0 | 0 | 0 | 1 | 1 | 1 |
| 001 | 0 | 0 | 0 | 1 | 1 | 1 | 1 | 0 |
| 010 | 0 | 0 | 1 | 0 | 1 | 1 | 0 | 1 |
| 011 | 0 | 1 | 0 | 0 | 1 | 0 | 1 | 1 |
| 100 | 0 | 1 | 0 | 0 | 1 | 0 | 1 | 1 |
| 101 | 0 | 0 | 1 | 0 | 1 | 1 | 0 | 1 |
| 110 | 0 | 0 | 0 | 1 | 1 | 1 | 1 | 0 |
| 111 | 1 | 0 | 0 | 0 | 0 | 1 | 1 | 1 |
| D(f) | 3 | 3 | 3 | 3 | 3 | 3 | 3 | 2 |
| $Q_E(f)$ | **2** | **2** | **2** | **2** | **2** | **2** | **2** | **2** |

By using $PAIR\_EQUALITY_4$ function we obtained a set of 4-argument Boolean functions, we denote it with $QFunc4$, where for each function there is an exact QQA which computes it with 2 queries. Totally 24 different functions were obtained $|QFunc4| = 24$ and half of it is presented in table 4.

Table 4. Results of applying transformation methods for $PAIR\_EQUALITY_4$ algorithm

| X | PAIR EQUALITY | 2nd method | | | 3rd method + 2nd method $\sigma_{VarNum} = \begin{pmatrix} 1234 \\ 1324 \end{pmatrix}$ | | | | 3rd method + 2nd method $\sigma_{VarNum} = \begin{pmatrix} 1234 \\ 3124 \end{pmatrix}$ | | | |
|---|---|---|---|---|---|---|---|---|---|---|---|---|
| | $\begin{pmatrix}1\\0\\0\\0\end{pmatrix}$ | $\begin{pmatrix}0\\1\\0\\0\end{pmatrix}$ | $\begin{pmatrix}0\\0\\1\\0\end{pmatrix}$ | $\begin{pmatrix}0\\0\\0\\1\end{pmatrix}$ | $\begin{pmatrix}1\\0\\0\\0\end{pmatrix}$ | $\begin{pmatrix}0\\1\\0\\0\end{pmatrix}$ | $\begin{pmatrix}0\\0\\1\\0\end{pmatrix}$ | $\begin{pmatrix}0\\0\\0\\1\end{pmatrix}$ | $\begin{pmatrix}1\\0\\0\\0\end{pmatrix}$ | $\begin{pmatrix}0\\1\\0\\0\end{pmatrix}$ | $\begin{pmatrix}0\\0\\1\\0\end{pmatrix}$ | $\begin{pmatrix}0\\0\\0\\1\end{pmatrix}$ |
| 0000 | **1** | 0 | 0 | 0 | 0 | 0 | 0 | **1** | **1** | 0 | 0 | 0 |
| 0001 | 0 | **1** | 0 | 0 | 0 | **1** | 0 | 0 | 0 | **1** | 0 | 0 |
| 0010 | 0 | 0 | **1** | 0 | 0 | **1** | 0 | 0 | 0 | 0 | **1** | 0 |
| 0011 | 0 | 0 | 0 | **1** | 0 | 0 | 0 | **1** | 0 | 0 | 0 | **1** |
| 0100 | 0 | **1** | 0 | 0 | 0 | 0 | **1** | 0 | 0 | 0 | **1** | 0 |
| 0101 | **1** | 0 | 0 | 0 | **1** | 0 | 0 | 0 | 0 | 0 | 0 | **1** |
| 0110 | 0 | 0 | 0 | **1** | **1** | 0 | 0 | 0 | **1** | 0 | 0 | 0 |
| 0111 | 0 | 0 | **1** | 0 | 0 | 0 | **1** | 0 | 0 | **1** | 0 | 0 |
| 1000 | 0 | 0 | **1** | 0 | 0 | 0 | **1** | 0 | 0 | **1** | 0 | 0 |
| 1001 | 0 | 0 | 0 | **1** | **1** | 0 | 0 | 0 | **1** | 0 | 0 | 0 |
| 1010 | **1** | 0 | 0 | 0 | **1** | 0 | 0 | 0 | 0 | 0 | 0 | **1** |
| 1011 | 0 | **1** | 0 | 0 | 0 | 0 | **1** | 0 | 0 | 0 | **1** | 0 |
| 1100 | 0 | 0 | 0 | **1** | 0 | 0 | 0 | **1** | 0 | 0 | 0 | **1** |
| 1101 | 0 | 0 | **1** | 0 | 0 | **1** | 0 | 0 | 0 | 0 | **1** | 0 |
| 1110 | 0 | **1** | 0 | 0 | 0 | **1** | 0 | 0 | 0 | **1** | 0 | 0 |
| 1111 | **1** | 0 | 0 | 0 | 0 | 0 | 0 | **1** | **1** | 0 | 0 | 0 |
| D(f) | 4 | 4 | 4 | 4 | 4 | 4 | 4 | 4 | 4 | 4 | 4 | 4 |
| $Q_E(f)$ | **2** | **2** | **2** | **2** | **2** | **2** | **2** | **2** | **2** | **2** | **2** | **2** |

## 5 Algorithm Constructing Methods

In this section we will present several quantum query algorithm constructing methods. Each method requires explicitly specified exact QQAs on input, but as a result a bounded-error QQA for more complex function is constructed. Our methods maintain quantum query complexity for complex function in comparison to increased deterministic complexity, thus enlarging the gap between classical and quantum complexities of an algorithm. We offer a general constructions for computing *AND*, *OR* and *MAJORITY* kinds of Boolean functions.

### 5.1 Bounded-error QQA for 6 variable function

We consider composite Boolean function, where two instances of *EQUALITY$_3$* (section 3.1) are joined with logical AND operation:

$$EQUALITY_3^{\wedge 2}(x_1,...,x_6) = (\neg(x_1 \oplus x_2) \wedge \neg(x_2 \oplus x_3)) \wedge (\neg(x_4 \oplus x_5) \wedge \neg(x_5 \oplus x_6))$$

**Deterministic complexity.** $D(EQUALITY_3^{\wedge 2}) = 6$, by sensitivity on X=111111.

**Algorithm 3.** Our approach in designing an algorithm for $EQUALITY_3^{\wedge 2}$ is to employ quantum parallelism and superposition principle. We execute algorithm pattern defined by original algorithm for *EQUALITY$_3$* in parallel for both blocks of $EQUALITY_3^{\wedge 2}$ variables. Finally we apply additional quantum gate to correlate amplitude distribution. Algorithm flow is depicted explicitly in figure 3.

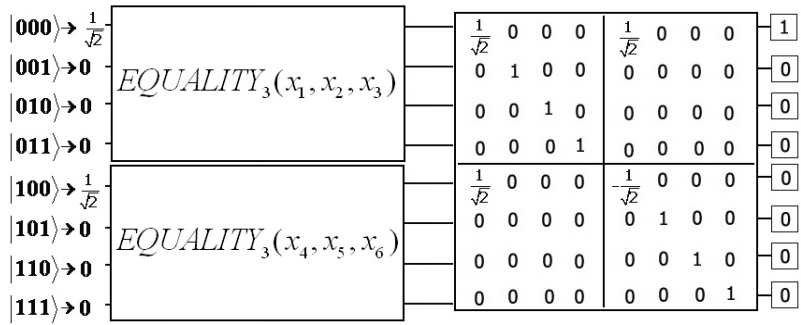

Fig. 3 Bounded-error QQA for $EQUALITY_3^{\wedge 2}$

**Quantum complexity.** Algorithm 3 computes $EQUALITY_3^{\wedge 2}$ using 2 queries with correct answer probability $p = 3/4$: $Q_{3/4}(EQUALITY_3^{\wedge 2}) = 2$.

**Proof.**
To calculate probabilities of obtaining correct function value it is enough to examine 4 cases depending on the value of each term of $EQUALITY_3^{\wedge 2}$. Results are presented in a table below. We use wildcards "?" and "*" to denote that exactly one value under the same wildcard is $\pm\frac{1}{\sqrt{2}}$ (we don't care which one), but all others are zeroes.

Table 5. Calculation of probabilities depending on algorithm flow for $EQUALITY_3^{\wedge 2}$.

| $EQUALITY_3$ $(x_1,x_2,x_3)$ | $EQUALITY_3$ $(x_4,x_5,x_6)$ | Amplitude distribution before last gate | Amplitude distribution after last gate | $p("1")$ |
|---|---|---|---|---|
| 0 | 0 | $(0,?,?,?,0,*,*,*)$ | $(0,?,?,?,0,*,*,*)$ | **0** |
| 0 | 1 | $\left(0,?,?,?,\frac{1}{\sqrt{2}},0,0,0\right)$ | $\left(\frac{1}{2},?,?,?,-\frac{1}{2},0,0,0\right)$ | **1/4** |
| 1 | 0 | $\left(\frac{1}{\sqrt{2}},0,0,0,0,?,?,?\right)$ | $\left(\frac{1}{2},0,0,0,\frac{1}{2},?,?,?\right)$ | **1/4** |
| 1 | 1 | $\left(\frac{1}{\sqrt{2}},0,0,0,\frac{1}{\sqrt{2}},0,0,0\right)$ | $(1,0,0,0,0,0,0,0)$ | **1** |

So, we have $p("1")=1$ and $p("0")=3/4$, we did not use additional queries, thus estimation $Q_{3/4}(EQUALITY_3^{\wedge 2})=2$ is proved. □

### 5.2 First Constructing Method – $AND(f_1,f_2)$

In this section we will generalize approach used in previous section. To be able to use generalized version of method we will limit ourselves to examining only exact QQA with specific properties.

**Property 2+** *We say that exact QQA satisfies Property2+ IFF there is exactly one accepting basic state and on any input for its amplitude $\alpha \in C$ only two values are possible before the final measurement: either $\alpha = 0$ or $\alpha = 1$.*

Algorithm 1 presented in section 3.1 satisfies Property 2+.

**Property 2-** *We say that exact QQA satisfies Property2- IFF there is exactly one accepting basic state and on any input for its amplitude $\alpha \in C$ only two values are possible before the final measurement: either $\alpha = 0$ or $\alpha = -1$.*

**Lemma 1.** *It is possible to transform an algorithm that satisfies Property2- to an algorithm that satisfies Property2+ by applying additional unitary transformation.*

**Proof.** Let's assume that we have QQA satisfying *Property2-* and *k* is the number of accepting output. To transform algorithm to satisfy *Property2+* apply the following quantum gate: $U = (u_{ij}) = \begin{cases} 0, & \text{if } i \neq j \\ 1, & \text{if } i = j \neq k \\ -1, & \text{if } i = j = k \end{cases}$

□

> **First constructing method - AND($f_1, f_2$)**
>
> **Input.**
> - Two exact QQAs A1 and A2 satisfying *Property2+* that compute correspondingly Boolean functions $f_1(X_1)$ and $f_2(X_2)$.
>
> **Transformation actions.**
>
> 1. If A1 and A2 utilize quantum systems of different size, then extend the smallest one with auxiliary space to obtain equal number of amplitudes. We denote the dimension of obtained Hilbert spaces with *m*.
> 2. For new algorithm utilize a quantum system with 2*m* amplitudes.
> 3. Combine unitary transformations and queries of A1 and A2 in the following way: $U_i = \begin{pmatrix} U_i^1 & O \\ O & U_i^2 \end{pmatrix}$, here *O*'s are $m \times m$ zero-matrices, $U_i^1$ and $U_i^2$ are either unitary transformations or query transformations of *A1* and *A2*.
> 4. Start computation from the state $\langle \psi | = (1/\sqrt{2}, 0,...,0, 1/\sqrt{2}, 0,...,0)$.
> 5. Before the final measurement apply additional unitary gate. Let's denote the positions of accepting outputs of A1 and A2 by $acc_1$ and $acc_2$. Then the final gate is defined as follows:
>
> $$U = (u_{ij}) = \begin{cases} 1, & \text{if } (i = j) \& (i \neq acc_1) \& (i \neq (m + acc_2)) \\ 1/\sqrt{2}, & \text{if } (i = j = acc_1) \text{ OR } (i = j = (m + acc_2)) \\ 1/\sqrt{2}, & \text{if } (i = acc_1) \& (j = (m + acc_2)) \text{ OR } (i = (m + acc_2)) \& (j = acc_1) \\ 0, & \text{otherwise} \end{cases}$$
>
> 6. Define as accepting output exactly one basic state $|acc_1\rangle$.
>
> **Output.** A bounded-error QQA *A* computing a function $F(X) = f_1(X_1) \wedge f_2(X_2)$ with probability $p = 3/4$ and complexity $Q_{3/4}(A) = \max(Q_E(A_1), Q_E(A_2))$.

Box 4. Description of the first constructing method for AND($f_1, f_2$)

### 5.3 Bounded-error quantum algorithm for 8 variable function

Next step is to realize similar approach for *OR* operation. This time we take second exact algorithm for *PAIR_EQUALITY$_4$* as a base.

We consider composite Boolean function, where two instances of *PAIR_EQUALITY$_4$* are joined with OR operation:

$$PAIR\_EQUALITY_4^{\vee 2}(x_1,...,x_8) = PAIR\_EQUALITY_4(x_1, x_2, x_3, x_4) \vee PAIR\_EQUALITY_4(x_5, x_6, x_7, x_8)$$

$$PAIR\_EQUALITY_4^{\vee 2}(x_1,...,x_8) = \left(\neg(x_1 \oplus x_2) \wedge \neg(x_3 \oplus x_4)\right) \vee \left(\neg(x_5 \oplus x_6) \wedge \neg(x_7 \oplus x_8)\right)$$

We succeeded in constructing quantum algorithm for $PAIR\_EQUALITY_4^{\vee 2}$, however algorithm structure is more complex than in *AND* operation case.

**Algorithm 4.** This time we use 4 qubit quantum system, so totally there are 16 amplitudes. First we execute $PAIR\_EQUALITY_4$ algorithm pattern in parallel on first 8 amplitudes, and then apply two additional quantum gates $U_{SWAP}$ and $U_{OR}$:

$$U_{SWAP} = \begin{pmatrix} 1 & 0 & 0 & 0 & 0 & 0 & 0 & 0 & 0 & 0 & .. & 0 \\ 0 & 0 & 0 & 0 & [1] & 0 & 0 & 0 & 0 & 0 & .. & 0 \\ 0 & 0 & 1 & 0 & 0 & 0 & 0 & 0 & 0 & 0 & .. & 0 \\ 0 & 0 & 0 & 1 & 0 & 0 & 0 & 0 & 0 & 0 & .. & 0 \\ 0 & [1] & 0 & 0 & 0 & 0 & 0 & 0 & 0 & 0 & .. & 0 \\ 0 & 0 & 0 & 0 & 0 & 0 & 0 & [1] & 0 & 0 & .. & 0 \\ 0 & 0 & 0 & 0 & 0 & 0 & 1 & 0 & 0 & 0 & .. & 0 \\ 0 & 0 & 0 & 0 & 0 & 0 & 0 & 1 & 0 & 0 & .. & 0 \\ 0 & 0 & 0 & 0 & [1] & 0 & 0 & 0 & 0 & 0 & .. & 0 \\ 0 & 0 & 0 & 0 & 0 & 0 & 0 & 0 & 0 & 1 & .. & 0 \\ .. & .. & .. & .. & .. & .. & .. & .. & .. & .. & .. & .. \\ 0 & 0 & 0 & 0 & 0 & 0 & 0 & 0 & 0 & 0 & .. & 1 \end{pmatrix}$$

$$U_{OR} = \begin{pmatrix} \frac{1}{\sqrt{2}} & \frac{1}{\sqrt{2}} & 0 & 0 & 0 & 0 & 0 & 0 & 0 & 0 & 0 & 0 & 0 & 0 & 0 & 0 \\ \frac{1}{\sqrt{2}} & -\frac{1}{\sqrt{2}} & 0 & 0 & 0 & 0 & 0 & 0 & 0 & 0 & 0 & 0 & 0 & 0 & 0 & 0 \\ 0 & 0 & 1/2 & 1/2 & 1/2 & 1/2 & 0 & 0 & 0 & 0 & 0 & 0 & 0 & 0 & 0 & 0 \\ 0 & 0 & 1/2 & -1/2 & 1/2 & -1/2 & 0 & 0 & 0 & 0 & 0 & 0 & 0 & 0 & 0 & 0 \\ 0 & 0 & 1/2 & 1/2 & -1/2 & -1/2 & 0 & 0 & 0 & 0 & 0 & 0 & 0 & 0 & 0 & 0 \\ 0 & 0 & 1/2 & -1/2 & -1/2 & 1/2 & 0 & 0 & 0 & 0 & 0 & 0 & 0 & 0 & 0 & 0 \\ 0 & 0 & 0 & 0 & 0 & 0 & 1/2 & 1/2 & 1/2 & 1/2 & 0 & 0 & 0 & 0 & 0 & 0 \\ 0 & 0 & 0 & 0 & 0 & 0 & 1/2 & -1/2 & 1/2 & -1/2 & 0 & 0 & 0 & 0 & 0 & 0 \\ 0 & 0 & 0 & 0 & 0 & 0 & 1/2 & 1/2 & -1/2 & -1/2 & 0 & 0 & 0 & 0 & 0 & 0 \\ 0 & 0 & 0 & 0 & 0 & 0 & 1/2 & -1/2 & -1/2 & 1/2 & 0 & 0 & 0 & 0 & 0 & 0 \\ 0 & 0 & 0 & 0 & 0 & 0 & 0 & 0 & 0 & 0 & 1 & 0 & 0 & 0 & 0 & 0 \\ 0 & 0 & 0 & 0 & 0 & 0 & 0 & 0 & 0 & 0 & 0 & 1 & 0 & 0 & 0 & 0 \\ 0 & 0 & 0 & 0 & 0 & 0 & 0 & 0 & 0 & 0 & 0 & 0 & 1 & 0 & 0 & 0 \\ 0 & 0 & 0 & 0 & 0 & 0 & 0 & 0 & 0 & 0 & 0 & 0 & 0 & 1 & 0 & 0 \\ 0 & 0 & 0 & 0 & 0 & 0 & 0 & 0 & 0 & 0 & 0 & 0 & 0 & 0 & 1 & 0 \\ 0 & 0 & 0 & 0 & 0 & 0 & 0 & 0 & 0 & 0 & 0 & 0 & 0 & 0 & 0 & 1 \end{pmatrix}$$

Quantum measurement:
$$QM = ([1,1], [1,0,0,0], [1,0,0,0], 0, 0, 0, 0, 0, 0)$$

Complete algorithm structure is presented in figure 4.

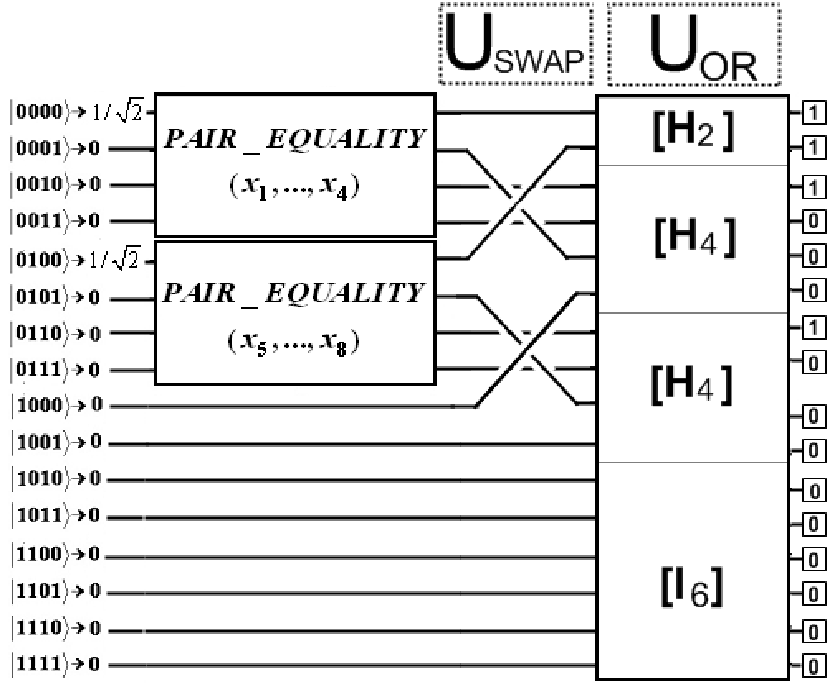

Fig. 4 Bounded-error QQA for $PAIR\_EQUALITY_4^{\vee 2}$

**Quantum complexity.** Algorithm 4 computes $PAIR\_EQUALITY_4^{\vee 2}$ using 2 queries with correct answer probability $p = 5/8$: $Q_{5/8}(PAIR\_EQUALITY_4^{\vee 2}) = 2$.

**Proof.** We demonstrate computation process results, what cover all possible inputs.

**I.** $PAIR\_EQUALITY_4(x_1, x_2, x_3, x_4) = 1$ and $PAIR\_EQUALITY_4(x_5, x_6, x_7, x_8) = 1$

| Amplitude distribution before $U_{OR}$ | Amplitude distribution before measurement | $p("1")$ |
|---|---|---|
| $\left(\left[\pm\frac{1}{\sqrt{2}}, \pm\frac{1}{\sqrt{2}}\right], [0,0,0,0], [0,0,0,0], 0,0,0,0,0,0\right)$ | $\pm(1,0,0,0,0,0,0,0,0,0,0,0,0,0,0,0)$ or $\pm(0,1,0,0,0,0,0,0,0,0,0,0,0,0,0,0)$ | 1 |

**II.** $PAIR\_EQUALITY_4(x_1, x_2, x_3, x_4) = 1$ and $PAIR\_EQUALITY_4(x_5, x_6, x_7, x_8) = 0$

| Amplitude distribution before $U_{OR}$ | Amplitude distribution before measurement | $p("1")$ |
|---|---|---|
| $\left(\left[\pm\frac{1}{\sqrt{2}}, 0\right], [0,0,0,0], [?,?,?,0],\right.$ $\left.0,0,0,0,0,0\right)$ | $\left(\left[\pm\frac{1}{2}, \pm\frac{1}{2}\right], [0,0,0,0],\right.$ $\left.\left[\pm\frac{1}{2\sqrt{2}}, \pm\frac{1}{2\sqrt{2}}, \pm\frac{1}{2\sqrt{2}}, \pm\frac{1}{2\sqrt{2}}\right], 0,0,0,0,0,0\right)$ | $\frac{1}{4} + \frac{1}{4} + \frac{1}{8} =$ $= \frac{5}{8}$ |

**III.** $PAIR\_EQUALITY_4(x_1, x_2, x_3, x_4) = 0$ and $PAIR\_EQUALITY_4(x_5, x_6, x_7, x_8) = 1$

| Amplitude distribution before $U_{OR}$ | Amplitude distribution before measurement | $p("1")$ |
|---|---|---|
| $\left(\left[0, \pm\frac{1}{\sqrt{2}}\right], [?,?,?,0], [0,0,0,0],\right.$ $\left.0,0,0,0,0,0\right)$ | $\left(\left[\pm\frac{1}{2}, \pm\frac{1}{2}\right], \left[\pm\frac{1}{2\sqrt{2}}, \pm\frac{1}{2\sqrt{2}}, \pm\frac{1}{2\sqrt{2}}, \pm\frac{1}{2\sqrt{2}}\right],\right.$ $\left.[0,0,0,0], 0,0,0,0,0,0\right)$ | $\frac{1}{4} + \frac{1}{4} + \frac{1}{8} =$ $= \frac{5}{8}$ |

**IV.** $PAIR\_EQUALITY_4(x_1, x_2, x_3, x_4) = 0$ and $PAIR\_EQUALITY_4(x_5, x_6, x_7, x_8) = 0$

| Amplitude distribution before $U_{OR}$ | Amplitude distribution before measurement | $p("1")$ |
|---|---|---|
| $\left([0,0], [?,?,?,0], [*,*,*,0],\right.$ $\left.0,0,0,0,0,0\right)$ | $\left([0,0], \left[\pm\frac{1}{2\sqrt{2}}, \pm\frac{1}{2\sqrt{2}}, \pm\frac{1}{2\sqrt{2}}, \pm\frac{1}{2\sqrt{2}}\right],\right.$ $\left.\left[\pm\frac{1}{2\sqrt{2}}, \pm\frac{1}{2\sqrt{2}}, \pm\frac{1}{2\sqrt{2}}, \pm\frac{1}{2\sqrt{2}}\right], 0,0,0,0,0,0\right)$ | $\frac{1}{8} + \frac{1}{8} = \frac{1}{4}$ |

Correct function result is always obtained with probability not less than 5/8, thus complexity estimation is proved.

□

### 5.4 Second Constructing Method – $OR(f_1, f_2)$

In this section we generalize approach for computing composite Boolean functions matching $OR(f_1, f_2)$ pattern.

First, we define next QQA property.

**Property 3** *We say that exact QQA satisfies Property3 IFF*
- *it satisfies Property1;*
- *there is exactly one accepting basic state;*
- *on any input accepting state amplitude value before measurement is $\alpha \in \{-1, 0, 1\}$*

Algorithm 1 and Algorithm 2 from section 3 both satisfy Property3.

The following lemma will be useful during method application.

**Lemma 2.** *For any QQA on any computation step it is possible to swap amplitude values in arbitrary order by applying specific quantum gate.*

**Proof.** Assume we need to swap amplitude values according to permutation $\sigma = \begin{pmatrix} \alpha_1 & \alpha_2 & ... & \alpha_n \\ \beta_1 & \beta_2 & ... & \beta_n \end{pmatrix}$. Then we can define quantum gate $U_{SWAP} = \{u_{ij}\}$ elements as:

- $\forall k \in \{1...n\}: u_{\alpha_k \beta_k} = 1$;
- $u_{ij} = 0$, in all other cases. □

Now we are ready to formulate a method for computing $OR(f_1, f_2)$ kind of functions. For simplicity we consider only such input algorithms, which employ 2 qubit system. However, approach can be generalized for quantum systems of arbitrary size.

---

***Second constructing method – $OR(f_1, f_2)$***

**Input.**
- Two exact QQAs A1 and A2 satisfying *Property3*, which use quantum systems with 2 qubits and compute correspondingly Boolean functions $f_1(X_1)$ and $f_2(X_2)$.

**Transformation actions.**

1. Use 4 qubit quantum system for new algorithm, totally $2^4 = 16$ basic states.

2. Convert initial state $\langle \vec{0} | = (1, 0, 0, 0, ..., 0)$ into state:

$$\langle \psi | = \left( \left[ \frac{1}{\sqrt{2}}, 0, 0, 0 \right], \left[ \frac{1}{\sqrt{2}}, 0, 0, 0 \right], 0, 0, 0, 0, 0, 0, 0, 0 \right)$$

3. Combine A1 and A2 unitary and query transformations in the following way:

$$U_i = \begin{pmatrix} \left[ U_i^1 \right] & O_{4x4} & O_{4x8} \\ O_{4x4} & \left[ U_i^2 \right] & O_{4x8} \\ O_{8x4} & O_{8x4} & \left[ I_8 \right] \end{pmatrix}, \text{ where } [I_8] \text{ is 8x8 identity matrix.}$$

4. Apply amplitude swapping gate $U_{SWAP}$, which was defined in the proof of lemma 2, to arrange amplitudes in the following order:

- $1^{st}$ amplitude $\leftrightarrow$ first sub-algorithm accepting amplitude;
- $2^{nd}$ amplitude $\leftrightarrow$ second sub-algorithm accepting amplitude;
- 3, 4, 5 amplitudes $\leftrightarrow$ first sub-algorithm rejecting amplitudes;
- 7, 8, 9 amplitudes $\leftrightarrow$ second sub-algorithm rejecting amplitudes.

5. Apply the last quantum gate, which was precisely defined in previous section:

$$U_{OR} = \begin{pmatrix} [H_2] & O_{2x4} & O_{2x4} & O_{2x6} \\ O_{4x2} & [H_4] & O_{4x4} & O_{4x6} \\ O_{4x2} & O_{4x4} & [H_4] & O_{4x6} \\ O_{6x2} & O_{6x4} & O_{6x4} & [I_6] \end{pmatrix}$$

6. Assign function values to algorithm outputs s follows:
$$QM = ([1,1],[1,0,0,0],[1,0,0,0],0,0,0,0,0,0)$$

**Output.** A bounded-error QQA $A$ computing a function $F(X) = f_1(X_1) \vee f_2(X_2)$ with probability $p = 5/8$ and complexity $Q_{5/8}(A) = \max(Q_E(A_1), Q_E(A_2))$.

Box 5. Description of the second constructing method for $OR(f_1, f_2)$

### 5.5 Bounded-error quantum algorithm for 12 variable function

Let us try to increase the effect gained by employing quantum parallelism. Next idea is to execute 4 instances of algorithm in parallel, adjusting algorithm parameters in appropriate way. We will take as a pattern function $EQUALITY_3$ from section 3.1.

Designed algorithm and additional gates are presented in figure 5 and below. Algorithm computes some 12-variable Boolean function with bounded-error.

**Algorithm 5.**

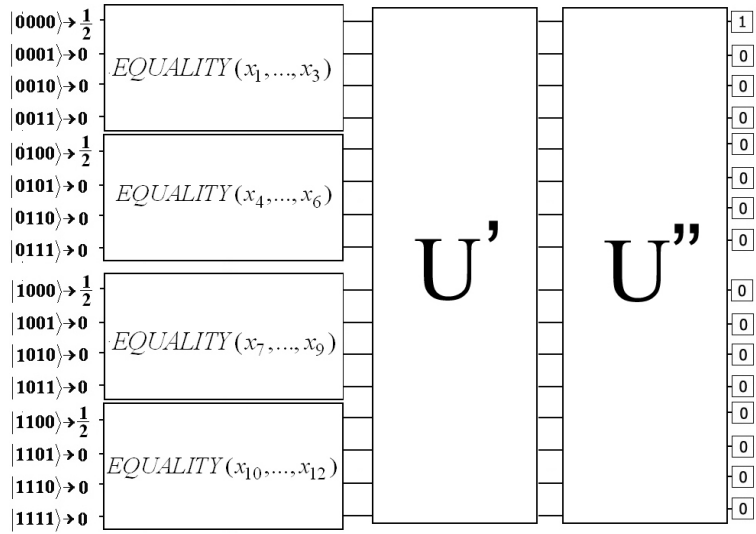

Fig. 5 Bounded-error quantum query algorithm for 12-variable function

Additional quantum gates (empty matrix cells correspond to "0"):

$$U' = \begin{pmatrix} \frac{1}{\sqrt{2}} & 0 & 0 & 0 & \frac{1}{\sqrt{2}} & 0 & 0 & 0 & 0 & 0 & 0 & 0 & 0 & 0 & 0 & 0 \\ 0 & 1 & 0 & & & & & 0 & & & & & 0 & & & \\ 0 & & 1 & 0 & & & & 0 & & & & & 0 & & & \\ 0 & & & 1 & 0 & & & 0 & & & & & 0 & & & \\ \frac{1}{\sqrt{2}} & 0 & 0 & 0 & \frac{-1}{\sqrt{2}} & & & 0 & & & & & 0 & & & \\ 0 & & & & & 1 & & 0 & & & & & 0 & & & \\ 0 & & & & & & 1 & 0 & & & & & 0 & & & \\ 0 & & & & & & & 1 & 0 & & & & 0 & & & \\ 0 & 0 & 0 & 0 & & 0 & 0 & 0 & \frac{1}{\sqrt{2}} & 0 & 0 & 0 & \frac{1}{\sqrt{2}} & 0 & 0 & 0 \\ 0 & & & & & & & & 0 & 1 & & & 0 & & & \\ 0 & & & & & & & & 0 & & 1 & & 0 & & & \\ 0 & & & & & & & & 0 & & & 1 & 0 & & & \\ 0 & & & & & & & & \frac{1}{\sqrt{2}} & 0 & 0 & 0 & \frac{-1}{\sqrt{2}} & & & \\ 0 & & & & & & & & 0 & & & & & 1 & & \\ 0 & & & & & & & & 0 & & & & & & 1 & \\ 0 & 0 & 0 & 0 & & 0 & 0 & 0 & 0 & 0 & 0 & 0 & & & & 1 \end{pmatrix}$$

$$U'' = \begin{pmatrix} \frac{1}{\sqrt{2}} & 0 & 0 & 0 & 0 & 0 & 0 & \frac{1}{\sqrt{2}} & 0 & 0 & 0 & 0 & 0 & 0 & 0 & 0 \\ 0 & 1 & & & & & & 0 & & & & & & & & 0 \\ 0 & & 1 & & & & & 0 & & & & & & & & 0 \\ 0 & & & 1 & & & & 0 & & & & & & & & 0 \\ 0 & & & & 1 & & & 0 & & & & & & & & 0 \\ 0 & & & & & 1 & & 0 & & & & & & & & 0 \\ 0 & & & & & & 1 & 0 & & & & & & & & 0 \\ \frac{1}{\sqrt{2}} & 0 & 0 & 0 & 0 & 0 & 0 & \frac{-1}{\sqrt{2}} & 0 & 0 & 0 & 0 & 0 & 0 & 0 & 0 \\ 0 & & & & & & & 0 & 1 & & & & & & & 0 \\ 0 & & & & & & & 0 & & 1 & & & & & & 0 \\ 0 & & & & & & & 0 & & & 1 & & & & & 0 \\ 0 & & & & & & & 0 & & & & 1 & & & & 0 \\ 0 & & & & & & & 0 & & & & & 1 & & & 0 \\ 0 & & & & & & & 0 & & & & & & 1 & & 0 \\ 0 & & & & & & & 0 & & & & & & & 1 & 0 \\ 0 & 0 & 0 & 0 & 0 & 0 & 0 & 0 & 0 & 0 & 0 & 0 & 0 & 0 & 0 & 1 \end{pmatrix}$$

After examination of algorithm computational flow and calculation of probabilities we obtained result that is formulated in the next statement.

**Quantum complexity.** Algorithm 5 computes function defined as:

$$F(x_1,...,x_{12}) = 1 \Leftrightarrow \begin{pmatrix} \text{Not less than 3 functions from: } EQUALITY(x_1,...,x_3), \\ EQUALITY(x_4,...,x_6), EQUALITY(x_7,...,x_9), \\ EQUALITY(x_{10},...,x_{12}) \text{ give value "1"}. \end{pmatrix}$$

and complexity is $Q_{9/16}(\text{Algorithm5}) = 2$.

**Deterministic complexity.** This time we did not achieve maximal possible gap. From the definition of function $F$ we find that sensitivity is $s(F) = 9$, thus in this case we can only register a gap $D(f) \geq 9$ vs. $Q_{9/16}(f)=2$.

### 5.6 Third Constructing Method - *MAJORITY*

We examined the structure of algorithm from previous section 5.5 and concluded that such approach would be useful for computing Boolean functions that belong to *MAJORITY* class.

**Definition 1.** *Boolean function MAJORITY$_n$(X), with $n = 2k+1$, $k \in N$ arguments is defined as:*

$$MAJORITY_{2k+1}(X) = 1 \Leftrightarrow \sum_{i=1}^{2k+1} x_i > k$$

When number of arguments is odd, then there always is clear majority of "0" or "1" in input vector. When number of function arguments is even, then the case when number of "0" and "1" is equal is not defined. We define another one class of Boolean functions for the case when number of function arguments is even.

**Definition 2.** *Boolean function MAJORITY_EVEN$_n$(X), with $n = 2k$, $k \in N$, $k > 0$ arguments is defined as:*

$$MAJORITY\_EVEN_{2k}(X) = 1 \Leftrightarrow \sum_{i=1}^{2k} x_i > k$$

So, when number of "0" and "1" in input vector is equal, then function value is "0".

In addition to *MAJORITY* function we define also *MAJORITY* composite construction. The difference is that in *MAJORITY* construction we use other Boolean functions as *MAJORITY* arguments.

**Definition 3.** *We define MAJORITY$_n$ construction ( $n = 2k+1$, $k \in N$ ) as a Boolean function where arguments are arbitrary Boolean functions $f_i$ and which is defined as:*

$$\left( MAJORITY_{2k+1}[f_1, f_2, ..., f_{2k+1}](X) = 1 \Leftrightarrow \sum_{i=1}^{2k+1} f_i(x_i) > k \right),$$

where $X = x_1 x_2 ... x_{2k+1}$

Construction *MAJORITY_EVEN$_n$* is defined in a similar way.

Let's again consider quantum algorithm 5 from the section 5.5. Definition of Boolean function was:

$$F(x_1, ..., x_{12}) = 1 \Leftrightarrow \begin{pmatrix} \text{Not less than 3 functions from: } EQUALITY(x_1, ..., x_3), \\ EQUALITY(x_4, ..., x_6), EQUALITY(x_7, ..., x_9), \\ EQUALITY(x_{10}, ..., x_{12}) \text{ give value "1".} \end{pmatrix}$$

Now we can rewrite it as:

$F(x_1, ..., x_{12}) = MAJORITY\_EVEN_4[EQUALITY_3](x_1, ..., x_{12}) = MAJORITY\_EVEN_4($
$EQUALITY_3(x_1, ..., x_3), EQUALITY_3(x_4, ..., x_6), EQUALITY_3(x_7, ..., x_9), EQUALITY_3(x_{10}, ..., x_{12}))$

Next we formulate a general algorithm constructing method for computing *MAJORITY_EVEN$_4$* construction.

| *Third constructing method - MAJORITY* |
|---|
| **Input.** <br> • Four exact QQAs A1, A2, A3, A4 satisfying *Property2+* that compute correspondingly Boolean functions $f_1(X_1), f_2(X_2), f_3(X_3), f_4(X_4)$. <br> **Transformation actions.** <br> 1. If any of input algorithms satisfy Property2-, then transform it to algorithm which satisfies Property2+ by applying lemma 1. <br> 2. Combine unitary and query transformations of input algorithms in the following |

way: $U_i = \begin{pmatrix} U_i^1 & O & O & O \\ O & U_i^2 & O & O \\ O & O & U_i^3 & O \\ O & O & O & U_i^4 \end{pmatrix}$, where $U_i^k$ is $k$-th algorithm transformation.

  $O$'s are zero sub-matrices, size depends on number of input algorithm amplitudes.

3. Start computation in a state:

$$\langle \psi | = \left( \frac{1}{2}, 0, ..., 0, \frac{1}{2}, 0, ..., 0, \frac{1}{2}, 0, ..., 0, \frac{1}{2}, 0, ..., 0 \right)$$

  where positions of 1/2 correspond to positions of the first amplitude of input algorithms.

4. Before the measurement apply two additional quantum transformations. We denote input algorithm accepting amplitude numbers as $\alpha_1$, $\alpha_2$, $\alpha_3$ and $\alpha_4$.

$$U' = \{u_{ij}\} = \begin{cases} 1, & \text{if } (i = j \neq \alpha_1) \vee (i = j \neq \alpha_2) \vee (i = j \neq \alpha_3) \vee (i = j \neq \alpha_4) \\ 1/\sqrt{2}, & \text{if } (i = j = \alpha_1) \vee (i = j = \alpha_3) \\ -1/\sqrt{2}, & \text{if } (i = j = \alpha_2) \vee (i = j = \alpha_4) \\ 1/\sqrt{2}, & \text{if } (i = \alpha_1 \,\&\, j = \alpha_2) \vee (i = \alpha_2 \,\&\, j = \alpha_1) \\ 1/\sqrt{2}, & \text{if } (i = \alpha_3 \,\&\, j = \alpha_4) \vee (i = \alpha_4 \,\&\, j = \alpha_3) \\ 0, & \text{otherwise} \end{cases}$$

$$U'' = \{u_{ij}\} = \begin{cases} 1, & \text{if } (i = j \neq \alpha_1) \vee (i = j \neq \alpha_2) \vee (i = j \neq \alpha_3) \vee (i = j \neq \alpha_4) \\ 1/\sqrt{2}, & \text{if } (i = j = \alpha_1) \\ -1/\sqrt{2}, & \text{if } (i = j = \alpha_3) \\ 1/\sqrt{2}, & \text{if } (i = \alpha_1 \,\&\, j = \alpha_3) \vee (i = \alpha_3 \,\&\, j = \alpha_1) \\ 0, & \text{otherwise} \end{cases}$$

5. Define as accepting state exactly one basic state $|\alpha_1\rangle$, that correspond to algorithm A1 accepting state.

**Output.** A bounded-error QQA *A* computing construction $MAJORITY\_EVEN_4[f_1, f_2, f_3, f_4](X)$, where $X = X_1 X_2 X_3 X_4$ with probability $p = 9/16$ and complexity $Q_{9/16}(A) = \max(Q_E(A_1), Q_E(A_2), Q_E(A_3), Q_E(A_4))$.

Box 6. Description of the third constructing method for *MAJORITY_EVEN*$_4$

By using a constant function $f(x)=1$ as one of constructing method input algorithms it is possible to achieve that resulting algorithm computes:

$$MAJORITY\_EVEN_4(f_1,f_2,f_3,1) = MAJORITY_3(f_1,f_2,f_3)$$

## 6 Results of Applying Methods

We applied transformation and designing methods to two basic exact QQAs described in section 3. Totally we obtained 32 exact QQAs and 512 QQAs with bounded error. Each algorithm computes different Boolean function and uses only 2 queries. Results are summarized in table 6. Here $n$ is number of variables of computable function.

Table 6. Results of transformation and constructing methods application

| Basic exact quantum algorithms | | | | |
|---|---|---|---|---|
| Set | Size | Number of arguments | Number of questions | Probability |
| *QFunc3* | **8** | 3 | 2 | 1 |
| *QFunc4* | **24** | 4 | 2 | 1 |
| Constructed algorithms sets | | | | |
| Set | Size | Number of arguments | Number of questions | Probability |
| *QFunc_AND* | **16** | 6 | 2 | 3/4 |
| *QFunc_OR* | **256** | 6,7,8 | 2 | 5/8 |
| *QFunc_MAJ_EVEN$_4$* | **256** | 12 | 2 | 9/16 |
| *QFunc_MAJORITY$_3$* | **64** | 9 | 2 | 9/16 |
| **Total** | **832** | | | |

The important point is that invention of each brand-new exact QQA with required properties will at once significantly increase a set of efficiently computable functions.

## 7 Conclusion

In this work we consider quantum query algorithm constructing problems. We have tried to develop some general approaches for designing algorithms for computing Boolean functions defined by logical formula. Main goal of research is to develop a framework for building ad-hoc quantum algorithms for arbitrary Boolean functions. In this paper we describe general constructions for designing quantum algorithms for *AND*, *OR* and *MAJORITY* kinds of Boolean functions.

First we presented two exact quantum query algorithms for 3 and 4 argument functions. Both algorithms save questions comparing to the best possible classical algorithm. Algorithms are used in further sections as a base for algorithm transformation and constructing methods.

Next we proposed techniques that allow transformation of an existing quantum query algorithm for a certain Boolean function so that the resulting algorithm computes a function with other logical structure. We illustrated methods by applying them to two basic exact algorithms.

Finally, we suggested approaches that allow building bounded-error quantum query algorithms for complex functions based on already known exact algorithms. Constructing methods include efficient solutions for *AND*, *OR* and *MAJORITY* constructions.

Combination of these three aspects allowed us to construct large sets of efficient quantum algorithms for various Boolean functions.

Further work in that direction could be to invent new efficient quantum algorithms that exceed already known separation from classical algorithms. Another important direction is improvement of general algorithm designing techniques.

## 8  Acknowledgments

I would like to thank my supervisor Rusins Freivalds for introducing me with quantum computation and for permanent support.

This research is supported by the European Social Fund.

## References

1.  H. Buhrman and R. de Wolf: "Complexity Measures and Decision Tree Complexity: A Survey". Theoretical Computer Science, v. 288(1): 21–43 (2002).

2.  R. de Wolf: "Quantum Computing and Communication Complexity". University of Amsterdam (2001).

3.  R. Cleve, A. Ekert, C. Macchiavello, et al. "Quantum Algorithms Revisited". Proceedings of the Royal Society, London, A454 (1998).

4.  J. Gruska: "Quantum Computing". McGraw-Hill (1999).

5.  M. Nielsen, I. Chuang: "Quantum Computation and Quantum Information". Cambridge University Press (2000).

6.  Ambainis: "Quantum query algorithms and lower bounds (survey article)". Proceedings of FOTFS III, to appear.

7.  Ambainis and R. de Wolf: "Average-case quantum query complexity". Journal of Physics A 34, pp 6741–6754 (2001).

8.  Ambainis. "Polynomial degree vs. quantum query complexity". Journal of Computer and System Sciences 72, pp. 220–238 (2006).